# A Multilayer Eigen-Sensitivity Method Using Loop Gain Model for Oscillation Diagnosis of Converter-Based System

Haoxiang Zong, *Student Member, IEEE*, Chen Zhang, *Member, IEEE*, Xu Cai, and Marta Molinas, *Member, IEEE*

*Abstract*—Loop gain-based eigen-sensitivity (LGES) is a useful frequency-domain tool for oscillation diagnosis of converter-based system. However, the existing theory is still scant in two aspects: participation factor (PF) is bound up with the frequency-domain modal characteristic that does not necessarily point to the stability as that of the time-domain eigen-sensitivity (i.e., PF of oscillation mode); a systematic LGES analysis framework containing both component- and parameter- level sensitivity is missing. These two factors hinder the application of LGES method on the proper evaluation of stability effects, which are closely related with the time-domain oscillation mode. To address these issues, this paper proposes a multilayer LGES method directed to the oscillation mode, and a full set of indices like PF, component and parameter sensitivity are established. The link from the eigen-sensitivity of frequency domain to that of time domain is revealed, through which it is shown how the proposed LGES method can facilitate the control parameter tuning-guided oscillation suppression. The effectiveness of the proposed LGES method is validated via case studies conducted on a generic AC/DC converter-based system.

*Index Terms*—converter, loop gain, eigen-sensitivity, oscillation diagnosis, participation factor, oscillation mode, suppression.

## I. INTRODUCTION

THE demand of low-carbon energy utilization stimulates the fast development of converters-interfaced renewable power generations (RPGs) like wind power, and converters-based grid interconnectors such as high-voltage dc (HVDC) transmission. This leads to a paradigm shift of traditional power system to a more converter-dominated power system [1]. Interactions among converter controls and electric networks result in a series of oscillation incidents, which are continually reported recently [2]. Thus, oscillatory stability issues of converter-based systems are becoming a very active research topic at present.

In this context, the impedance-based analysis method for the oscillatory stability analysis is widely applied [3], due to its superior traits of compactness and modular scalability. Early-stage researches are more focused on the mechanism analysis of those control-induced oscillation issues, and thus a single machine system [4], [5] (e.g., a single grid-tied converter) is usually adopted. As the advancement of research, current focus of analysis is gradually shifted to the multi-machine system [6], [7] (e.g., a multi-converters based system), concentrating more on oscillation diagnostics like oscillation source identification, propagation, attenuation, etc. This shift also greatly promotes the development of the impedance-based analysis method from the classical *single-source-load* method [3] (i.e., converter and grid respectively as the load and source subsystem) to the *multi-source-load* one [8],[9] (i.e., various converters and grids as a group of load and source subsystems). And for the latter case, both the source and load subsystems are modeled in the network form like impedance network or admittance network [9].

Given that the impedance/admittance network retains the structural information of the overall system [8], the network-based analysis has great potential of revealing multi-machine interactive behaviors and pinpointing crucial instability factors, through which a full view of the oscillation generation and propagation can be acquired [10]. In contrast, it is difficult to obtain such information via the (aggregated) single-source-load method, because the system structural information is lost by impedance aggregation [9]. One step further, to expose the structural information (e.g., nodes, infrastructures, etc.) in oscillation diagnostics, the eigen-sensitivity using the network-based frequency-domain model is developed. In this regard, depending on which type of frequency domain model is adopted, there are two representative ways [11]-[17]:

*The Nodal Admittance-based Eigen-Sensitivity (NAES)*: This method originates from the resonance modal analysis [11],[12] of passive transmission networks, and has been extended to the converter-based AC or AC/DC system[10], [13]-[15] recently. The main idea is to establish the system node admittance model $Y(s)$ in frequency domain, where passive elements and converters are modelled by two-port ac or three-port ac/dc admittances [13], [15]. Then, the oscillation mode is identified by calculating right-half-plane (RHP) zeros of $\det[Y(s)]$, based on which the eigenvalue decomposition is conducted to obtain the node sensitivity (referred to as the participation factor (PF)). And the derivative chain rule can be further applied to establish the component-level sensitivity beneath the PF and its next level sensitivity, i.e., parameter sensitivity [12]-[14]. Since

This work was supported by the National Natural Science Foundation of China under Grant 51837007. *(Corresponding author: Chen Zhang).*

Haoxiang Zong, Chen Zhang, and Xu Cai are with the Key Laboratory of Control of Power Transmission and Conversion of Ministry of Education, Shanghai Jiao Tong University (SJTU), 200240, Shanghai, China, also with the Department of Electrical Engineering, School of Electronic Information and Electrical Engineering, SJTU, 200240, Shanghai. (e-mails: {haoxiangzong, nealbc, xucai} @sjtu.edu.cn).

Marta Molinas is with Department of Engineering Cybernetics, Norwegian University of Science and Technology, Trondheim, 7491, Norway. (e-mail: marta.molinas@ntnu.no).

the above sensitivity indices are all with respect to the modal admittance, [16] also proposes a transformation method to connect them with the *time-domain* oscillation mode, so as to facilitate the controller tuning and oscillation suppression.

*The Loop Gain-based Eigen-Sensitivity (LGES)*: This method is inspired by the way of performing the generalized Nyquist criterion (GNC)-based stability test, for which a so-called *generalized source-load model* [17] is proposed. The whole system will be split into two subnetworks, one is the source subsystem with all passive elements and the other is the load subsystem with all converters. Then, the system loop gain $L(s)$ with multi-ports can be obtained, with which the same eigenvalue decomposition is applied to acquire loop gain-based PFs [18]. Compared with the NAES method, the existing LGES method is still scant in two aspects: a) the existing loop gain-based PF is a frequency-domain modal characteristic [18] that does not necessarily point to the stability as that of the time-domain eigen-sensitivity (i.e., PF of oscillation mode); b) a systematic analysis framework of LGES including the component and parameter -level sensitivity are missing.

To bridge the above gap and aimed for developing a multilayer LGES framework as the counterpart to the NAES method, this paper will contribute to the following aspects:
1) Propose a multilayer LGES method with respect to the time-domain oscillation mode, and indices like PF, component and parameter sensitivity are all established;
2) The link from eigen-sensitivity of frequency domain to that of time domain is revealed, through which it is shown how the proposed LGES method can facilitate the parameter tuning-based oscillation suppression.

The rest of paper is arranged as follows: Section II makes a review on the existing LGES method. Section III introduces the proposed multilayer LGES method. Section IV validates the correctness of the proposed LGES method. Section V shows the application of the proposed LGES method on the oscillation suppression. Section VI concludes the paper.

## II. INTRODUCTION OF EXISTING LGES ANALYSIS METHOD

This Section will briefly review the loop gain model and the existing calculation method of its eigen-sensitivity, whose parts can be improved are clearly pointed out.

### A. Loop Gain Model of Network-based Converter System

Recently, a frequency-domain loop gain model is proposed to characterize small-signal dynamics of multi-converters system in the network form, with the advantage of being immune to those nasty RHP-poles [15], [17]. As intuitively shown in Fig. 1, the whole system is split at the nodes (denoted as red circles) connected with converters. Then, generalized source and load subsystems [17] with multi-ports can be established, and the resulting closed-loop model is like:

$$V_{\text{node}}(s) = \frac{Z_{\text{net}}(s)}{I + Z_{\text{net}}(s)Y_{\text{con}}(s)} \cdot I_{\text{node}}(s) \quad (1)$$

where $I_{\text{node}}(s)$ and $V_{\text{node}}(s)$ are node currents and voltages, respectively. $Z_{\text{net}}(s)$ denotes the node impedance of the source subsystem with only passive elements, where internal nodes (denoted as blue circles) are eliminated via the Kron reduction. $Y_{\text{con}}(s)$ denotes the node admittance of the load subsystem containing only converters. $I$ is the identity matrix.

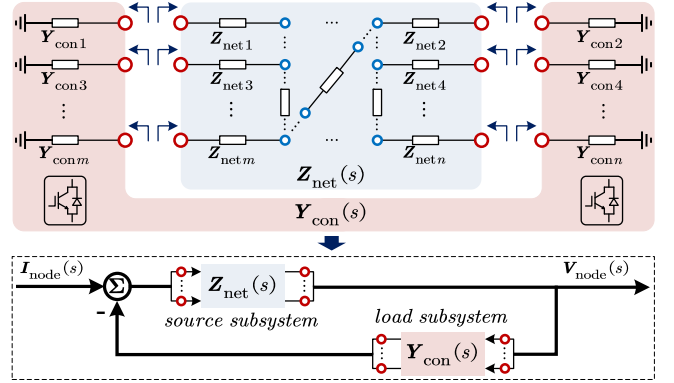

Fig. 1. Generalized source-load partition and the resulting closed-loop model.

Since $Z_{\text{net}}(s)$ is constituted of pure passive elements, it is free of RHP poles and zeros. Then, the stability of closed-loop system in (1) can be determined by its loop gain $L(s)$.

$$L(s) = I + Z_{\text{net}}(s)Y_{\text{con}}(s) \quad (2)$$

The system is stable if and only if there exists no RHP zero in $\det[L(s)]$. Supposing an unstable oscillation at $s=\lambda_0$ happens, i.e., an RHP zero appears, it should fulfill:

$$\det[L(\lambda_0)] = 0 \quad (3)$$

To simplify notation, $\lambda_0$ in brackets will be omitted hereinafter, i.e., $L(\lambda_0)$ will be denoted as the notation $L$.

### B. Loop Gain-based Participation Factor

The eigen-sensitivity analysis method [18] is also developed for such loop gain model, mainly the participation factor (PF), so as to facilitate the oscillation source locating. Specifically, the eigen-decomposition will be applied to $L$ as:

$$L = R^L \cdot \Lambda^L \cdot T^L$$
$$= \begin{bmatrix} r_1^L & r_2^L & \cdots & r_n^L \end{bmatrix} \cdot \begin{bmatrix} \Lambda_1^L & & & \\ & \Lambda_2^L & & \\ & & \ddots & \\ & & & \Lambda_n^L \end{bmatrix} \cdot \begin{bmatrix} t_1^L \\ t_2^L \\ \vdots \\ t_n^L \end{bmatrix} \quad (4)$$

where $\Lambda_i^L$ is defined as the modal loop gain; $R^L$ is the right column eigenvector, $T^L$ is the left row eigenvector and it has $R^L T^L = I$. The superscript '$L$' denotes that variables belong to the loop gain-based eigen-sensitivity (LGES), so as to distinguish from those variables of the node admittance-based eigen-sensitivity (NAES) [13].

The eigenvalue matrix $\Lambda^L$ can be represented as:

$$\Lambda^L = T^L \cdot L \cdot R^L \quad (5)$$

There will appear an eigenvalue $\Lambda_i^L$ in $\Lambda^L$ close to 0, which corresponds to that one Nyquist loci will cross the (-1, *j*0). This eigenvalue is defined as the critical modal loop gain. Supposing $\Lambda_1^L \approx 0$ (i.e., critical modal loop gain), it can be expressed as in (6) according to (5).

$$\Lambda_1^L = \boldsymbol{t}_1^L \cdot \boldsymbol{L} \cdot \boldsymbol{r}_1^L \tag{6}$$

where $\boldsymbol{t}_1^L$ and $\boldsymbol{r}_1^L$ are the corresponding critical eigenvectors.

Based on (6), loop gain-based PF (referred as *LG-PF* hereinafter) is established in [18] as:

$$\frac{\partial \Lambda_1^L}{\partial L_{ij}} = \boldsymbol{t}_1^L \cdot \frac{\partial \boldsymbol{L}}{\partial L_{ij}} \cdot \boldsymbol{r}_1^L = \left[\boldsymbol{r}_1^L \boldsymbol{t}_1^L\right]_{ji} = \left[\boldsymbol{PF}^L\right]_{ji} \tag{7}$$

where $\boldsymbol{PF}^L$ is defined as the *LG-PF matrix*.

It can be seen that the existing LG-PF reflects the sensitivity of system parts $L_{ij}$ to the critical modal loop gain $\Lambda_1^L$, which is in *frequency domain*. However, in terms of stability effects, the sensitivity of $L_{ij}$ to the *time-domain* oscillation mode $\lambda_0$ is more concerned. This is because the oscillation diagnosis and suppression are realized mainly through the movement of oscillation mode $\lambda_0$ from the unstable RHP to the stable left-half-plane (LHP). Also, the specific suppression strategy is conducted from the level of components and parameters, while their sensitivities are still absent.

### III. A Multilayer LGES Analysis Method with Respect to the Oscillation mode

To address above issues, this Section will introduce a multilayer LGES analysis method. Three sensitivity indices, including loop PF, component- and parameter- sensitivity, are established and all linked to the oscillation mode.

#### A. Outline of Multilayer LGES Index System

The oscillation mode $\lambda_0$, i.e., the RHP zero of $\det[\boldsymbol{L}(s)]$, is exactly the eigenvalue [14] of the state-space matrix $\boldsymbol{A}$ as shown in (8). In this sense, the oscillation mode $\lambda_0$ mainly reflects the system time-domain characteristics.

$$\lambda_0 \in \{\mathrm{eig}(\boldsymbol{A}) \Leftrightarrow \det[\boldsymbol{L}(s)] = 0\} \tag{8}$$

The critical modal loop gain $\Lambda_1^L$ is the eigenvalue of the frequency-domain loop gain matrix $\boldsymbol{L}$ as shown in (9). It mainly reflects the system frequency-domain characteristics.

$$\Lambda_1^L \in \{\mathrm{eig}(\boldsymbol{L})\} \tag{9}$$

The establishment procedure of the multilayer eigen-sensitivity is presented in Fig. 2. The first step is to bridge the gap between the critical modal loop gain $\Lambda_1^L$ and the oscillation mode $\lambda_0$, which will be derived concretely in Section B. With this relation, the existing PF of the critical modal loop gain can be transformed into that of the oscillation mode. Then, based on the derivative chain rule, the oscillation mode-directed component- and parameter- sensitivity can be computed, which will be presented carefully in Section C.

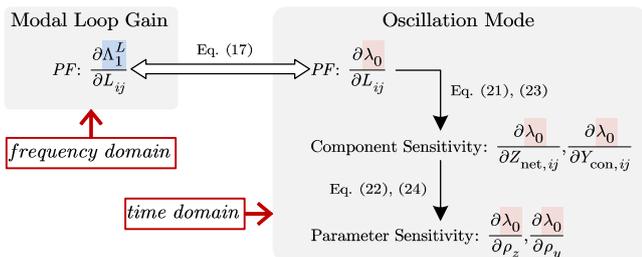

Fig. 2. Establishment of the multilayer eigen-sensitivity index system.

#### B. LG-PF of Oscillation Mode

As stated above, the connection between two kinds of PFs (i.e., $\frac{\partial \Lambda_1^L}{\partial L_{ij}}$ and $\frac{\partial \lambda_0}{\partial L_{ij}}$) will be derived first. Based on the derivative chain rule, $\det[\boldsymbol{L}]$ is utilized as a bridge to connect $\frac{\partial \Lambda_1^L}{\partial L_{ij}}$ with $\frac{\partial \lambda_0}{\partial L_{ij}}$.

$$\frac{\partial \lambda_0}{\partial L_{ij}} = \frac{\partial \lambda_0}{\partial \det[\boldsymbol{L}]} \cdot \frac{\partial \det[\boldsymbol{L}]}{\partial \Lambda_1^L} \cdot \frac{\partial \Lambda_1^L}{\partial L_{ij}} \tag{10}$$

Then, two partial derivatives concerning $\det[\boldsymbol{L}]$ needs to be calculated. Since there exists an RHP zero $\lambda_0$ (i.e., oscillation mode), $\det[\boldsymbol{L}(s)]$ can be represented in the pole-zero form as:

$$\det[\boldsymbol{L}(s)] = (s - \lambda_0) \cdot \underbrace{\frac{\prod_{i \neq 0}(s - \lambda_i)}{\prod(s - p_i)}}_{g^L(s)} \tag{11}$$

The partial derivative of $\lambda_0$ to $\det[\boldsymbol{L}(s)]$ can be obtained as:

$$\partial[\det(\boldsymbol{L}(s))] = -\partial \lambda_0 \cdot g^L(s) \tag{12}$$

By substituting $\lambda_0$ into $g^L(s)$, it has:

$$\frac{\partial \lambda_0}{\partial \det[\boldsymbol{L}]} = \left.\frac{\partial \lambda_0}{\partial[\det(\boldsymbol{L}(s))]}\right|_{s=\lambda_0} = -\left(g^L\right)^{-1} \tag{13}$$

where the notation $g^L$ represents the $g^L(\lambda_0)$.

Since the eigen-matrix has been derived in (4), $\det[\boldsymbol{L}(s)]$ can also be represented by the product of all eigenvalues as:

$$\det(\boldsymbol{L}(s)) = \Lambda_1^L(s) \cdot \prod_{i \neq 1}^n \Lambda_i^L(s) \tag{14}$$

Similar to (11)~(13), the partial derivative of $\det[\boldsymbol{L}]$ to $\Lambda_1^L$ can be obtained as in (15).

$$\frac{\partial \det[\boldsymbol{L}]}{\partial \Lambda_1^L} = \left.\frac{\partial[\det(\boldsymbol{L}(s))]}{\partial \Lambda_1^L(s)}\right|_{s=\lambda_0} = \prod_{i \neq 1}^n \Lambda_i^L \tag{15}$$

where the notation $\Lambda_i^L$ represents the $\Lambda_i^L(\lambda_0)$.

By substituting (13) and (15) into (10), the following relation can be acquired.

$$\frac{\partial \lambda_0}{\partial L_{ij}} = \underbrace{-\left(g^L\right)^{-1} \prod_{i \neq 1}^n \Lambda_i^L}_{\xi^L} \cdot \frac{\partial \Lambda_1^L}{\partial L_{ij}} = \xi^L \cdot \left[\boldsymbol{PF}^L\right]_{ji} \tag{16}$$

where $\xi^L$ is a complex-valued modification factor.

Based on (16), the LG-PF matrix with respect to the oscillation mode (denoted as $\boldsymbol{PF}_{\mathrm{mod}}^L$) can be expressed as:

$$\boldsymbol{PF}_{\mathrm{mod}}^L = \xi^L \cdot \boldsymbol{PF}^L \tag{17}$$

#### C. Component and Parameter Sensitivity of Oscillation Mode

1) *Components and Parameters Inside $\boldsymbol{Z}_{\mathrm{net}}$*

The sensitivity of impedance component (e.g., $Z_{\mathrm{net},ij}$) with respect to the oscillation mode can be calculated from that with respect to the critical modal loop gain.

$$\frac{\partial \lambda_0}{\partial Z_{\mathrm{net},ij}} = \frac{\partial \lambda_0}{\partial \det[\boldsymbol{L}]} \cdot \frac{\partial \det[\boldsymbol{L}]}{\partial \Lambda_1^L} \cdot \frac{\partial \Lambda_1^L}{\partial Z_{\mathrm{net},ij}} = \xi^L \cdot \frac{\partial \Lambda_1^L}{\partial Z_{\mathrm{net},ij}} \tag{18}$$

By substituting (6) into (18), it has

$$\frac{\partial \lambda_0}{\partial Z_{\text{net},ij}} = \xi^L \cdot \frac{\partial \Lambda_1^L}{\partial Z_{\text{net},ij}} = \xi^L \cdot \frac{\partial \left( t_1^L \left[ I + Z_{\text{net}} Y_{\text{con}} \right] r_1^L \right)}{\partial Z_{\text{net},ij}}$$
$$= \xi^L \cdot t_1^L \left( \frac{\partial I}{\partial Z_{\text{net},ij}} + \frac{\partial Z_{\text{net}}}{\partial Z_{\text{net},ij}} Y_{\text{con}} + Z_{\text{net}} \frac{\partial Y_{\text{con}}}{\partial Z_{\text{net},ij}} \right) r_1^L \quad (19)$$

$I$ and $Y_{\text{con}}$ do not contain any element concerning $Z_{\text{net},ij}$, meaning that their corresponding partial derivatives equal to zero. Then, (19) can be simplified as:

$$\frac{\partial \lambda_0}{\partial Z_{\text{net},ij}} = \xi^L \cdot t_1^L \left( \frac{\partial Z_{\text{net}}}{\partial Z_{\text{net},ij}} Y_{\text{con}} \right) r_1^L$$
$$= \xi^L \cdot t_1^L \begin{pmatrix} \begin{bmatrix} \ddots & 0 & \cdot\cdot \\ 0 & 1_{ij} & 0 \\ \cdot\cdot & 0 & \ddots \end{bmatrix} Y_{\text{con}} \end{pmatrix} r_1^L$$
$$= \xi^L \cdot \left( t_{1i}^L \cdot Y_{\text{con},j1} \cdot r_{11}^L + \cdots + t_{1i}^L \cdot Y_{\text{con},jn} \cdot r_{n1}^L \right) \quad (20)$$
$$= \xi^L \cdot \left[ Y_{\text{con},j1}, \ldots, Y_{\text{con},jn} \right] \cdot \left[ PF_{1i}^L, \ldots, PF_{ni}^L \right]^T$$
$$= \xi^L \cdot \left[ Y_{\text{con}} PF^L \right]_{ji}$$

By substituting (17) into (20), it can be obtained as:

$$\frac{\partial \lambda_0}{\partial Z_{\text{net},ij}} = \left[ Y_{\text{con}} PF_{\text{mod}}^L \right]_{ji} \quad (21)$$

Based on the derivative chain rule, the parameter sensitivity can be further derived from the above component sensitivity. Supposing parameter $\rho_z$ affects some impedance components $Z_{\text{net},ij}$, its sensitivity can be calculated as in (22).

$$\frac{\partial \lambda_0}{\partial \rho_z} = \sum \left( \frac{\partial \lambda_0}{\partial Z_{\text{net},ij}} \frac{\partial Z_{\text{net},ij}}{\partial \rho_z} \right) = \sum \left( \left[ Y_{\text{con}} PF_{\text{mod}}^L \right]_{ji} \frac{\partial Z_{\text{net},ij}}{\partial \rho_z} \right) \quad (22)$$

where $\Sigma$ represents that all components containing $\rho_z$ should be considered in the form of summation. The subscript '$z$' represents parameters of passive elements in $Z_{\text{net}}(s)$.

*2) Components and Parameters Inside $Y_{\text{con}}$*

The admittance component sensitivity can be calculated in a similar way as (20), where $Y_{\text{con},ij}$ is taken for derivation. At this time, $I$ and $Z_{\text{net}}$ will not contain any element concerning $Y_{\text{con},ij}$, (19) can be rewritten as:

$$\frac{\partial \lambda_0}{\partial Y_{\text{con},ij}} = \xi^L \cdot t_1^L \left( Z_{\text{net}} \frac{\partial Y_{\text{con}}}{\partial Y_{\text{con},ij}} \right) r_1^L$$
$$= \xi^L \cdot t_1^L \left( Z_{\text{net}} \begin{bmatrix} \ddots & 0 & \cdot\cdot \\ 0 & 1_{ij} & 0 \\ \cdot\cdot & 0 & \ddots \end{bmatrix} \right) r_1^L$$
$$= \xi^L \cdot \left( t_{11}^L \cdot Z_{\text{net},1i} \cdot r_{j1}^L + \cdots + t_{1n}^L \cdot Z_{\text{net},ni} \cdot r_{j1}^L \right) \quad (23)$$
$$= \xi^L \cdot \left[ PF_{j1}^L, \ldots, PF_{jn}^L \right] \cdot \left[ Z_{\text{net},1i}, \ldots, Z_{\text{net},ni} \right]^T$$
$$= \xi^L \cdot \left[ PF^L \cdot Z_{\text{net}} \right]_{ji} = \left[ PF_{\text{mod}}^L Z_{\text{net}} \right]_{ji}$$

TABLE I
SUMMARIZATION OF PROPOSED LGES ANALYSIS METHOD

| Eigen-Sensitivity Indices | | Calculation Formula |
|---|---|---|
| Participation Factor | | $\frac{\partial \lambda_0}{\partial L_{ij}} = \left[ PF_{\text{mod}}^L \right]_{ji}$ |
| Component Sensitivity | passive elements | $\frac{\partial \lambda_0}{\partial Z_{\text{net},ij}} = \left[ Y_{\text{con}} PF_{\text{mod}}^L \right]_{ji}$ |
| | converters | $\frac{\partial \lambda_0}{\partial Y_{\text{con},ij}} = \left[ PF_{\text{mod}}^L Z_{\text{net}} \right]_{ji}$ |
| Parameter Sensitivity | parameters of passive elements | $\frac{\partial \lambda_0}{\partial \rho_z} = \sum \left( \left[ Y_{\text{con}} PF_{\text{mod}}^L \right]_{ji} \frac{\partial Z_{\text{net},ij}}{\partial \rho_z} \right)$ |
| | parameters of converters | $\frac{\partial \lambda_0}{\partial \rho_y} = \sum \left( \left[ PF_{\text{mod}}^L Z_{\text{net}} \right]_{ji} \frac{\partial Y_{\text{con},ij}}{\partial \rho_y} \right)$ |

Supposing $\rho_y$ affects some admittance components $Y_{\text{con},ij}$, its sensitivity can be calculated similarly as in (24).

$$\frac{\partial \lambda_0}{\partial \rho_y} = \sum \left( \frac{\partial \lambda_0}{\partial Y_{\text{con},ij}} \frac{\partial Y_{\text{con},ij}}{\partial \rho_y} \right) = \sum \left( \left[ PF_{\text{mod}}^L Z_{\text{net}} \right]_{ji} \frac{\partial Y_{\text{con},ij}}{\partial \rho_y} \right) \quad (24)$$

where subscript '$y$' denotes converters' parameters in $Y_{\text{con}}(s)$.

So far, the multilayer eigen-sensitivity of the loop gain $L(s)$ has been established, and main sensitivity indices are summarized in Table I. These indices are all complex-valued involving information of both magnitude and phase to the oscillation mode, which can greatly facilitate the oscillation suppression. This will be explained carefully in Section V.

## IV. CASE STUDIES AND VALIDATION

This Section will implement some case studies on a typical AC/DC converter-dominated system, based on which the correctness of established multilayer LGES indices is verified.

### A. System Description

A three-terminal HVDC system shown in Fig. 3 is selected as the test case. Small-signal dynamics of the converter are modelled by a three-port ac/dc admittance model [15]. The model of the load subsystem $Y_{\text{con}}(s)$ is given in (25), and the source subsystem $Z_{\text{net}}(s)$ can be modelled in a similar way (more details can be referred to [17] and [19]).

$$Y_{\text{con}}(s) = blkdiag \left[ Y_{\text{con1}}(s), Y_{\text{con2}}(s), Y_{\text{con3}}(s) \right] \quad (25)$$

Notations of each non-zero $Z_{\text{net}}$ or $Y_{\text{con}}$ component and their internal parameters are listed in Table II. For this system, a typical marginal unstable condition is considered, i.e., increasing the PLL bandwidth $\alpha_{PQ}^{S1}$ of sending converter-2 (SEC-2). Fig. 4 presents the pole-zero distribution of this oscillation case, and there appears one pair of RHP zeros $\lambda_{1,2}=0.0794\pm j53.2$ (i.e., oscillation mode). The rest of poles and zeros will be used for calculating (11).

### B. Results of Multilayer LGES Analysis

*1) Results of LG-PF*

By substituting the obtained oscillation mode (e.g., $\lambda_1=0.0794+j53.2$) into $L(s)$, its eigenvalues can be calculated using (4) and results are listed in Table III. It can be seen that

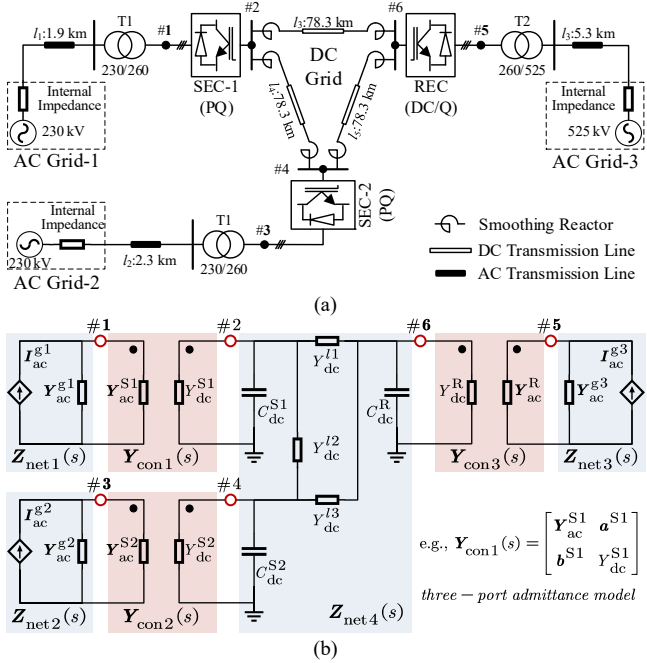

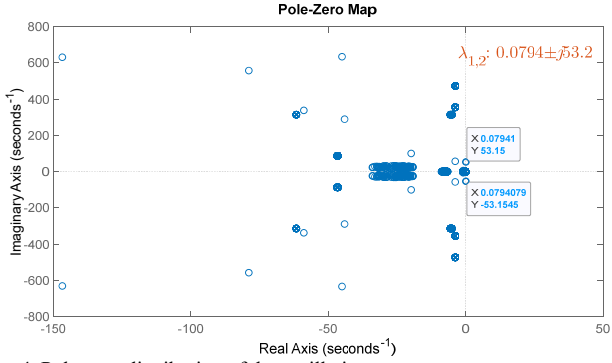

Fig. 3. Test system, (a) topology; (b) equivalent circuit.

Fig. 4. Pole-zero distribution of the oscillation case.

$\Lambda_2^L \approx 0$ is the critical modal loop gain, based on which $r_2^L$ and $t_2^L$ are extracted as critical eigenvectors. Then, according to (13), (15) and (16), both $g^L$ and $\prod_{i \neq 2}^{9} \Lambda_i^L$ are calculated to obtain the modification factor $\xi^L$ and results are given in Table III.

With the identified critical eigenvectors and modification factor, $\boldsymbol{PF}_{\text{mod}}^L$ can be obtained. To make data more readable, the trace of $|\boldsymbol{PF}_{\text{mod}}^L|$ is used as the per-unit base, and results of the normalized $\boldsymbol{PF}_{\text{mod}}^L$ are given in Fig. 5(a). It can be clearly seen that nodes #$3_d$, #$3_q$ are with the largest values, meaning that interactions between AC Grid-2 and SEC-2 are dominant in this oscillation case.

*2) Results of Component and Parameter Sensitivity*

With the calculation of $\boldsymbol{PF}_{\text{mod}}^L$, the component sensitivity concerning $\boldsymbol{Z}_{\text{net}}$ and $\boldsymbol{Y}_{\text{con}}$ can be obtained according to (21) and (23), whose results are given in Fig. 5(b) and (c). Similarly, $Z_{\text{net},44}$, $Z_{\text{net},45}$, $Z_{\text{net},54}$, $Z_{\text{net},55}$, $Y_{\text{con},44}$, $Y_{\text{con},45}$, $Y_{\text{con},54}$ and $Y_{\text{con},55}$ are with large values, meaning that the internal impedance of AC Grid-2 and ac-side of SEC-2 are the root cause of the system oscillation. The sensitivity of their internal parameters is also calculated and shown at the bottom of Fig. 5(b) and (c), respectively. Four parameters $L_{\text{ac}}^{g2}$, $\alpha_{cc}^{S1}$, $\alpha_{pll}^{S2}$ and $\alpha_{cc}^{S2}$ are with high values, which should be the focus of the parameter tuning for the oscillation suppression.

*C. Validation*

The above theoretical calculation results will be compared with actual tuning results, so as to justify their accuracies.

*1) Validation of LG-PF*

The actual tuning results for the above PF and sensitivity indices will be calculated here. For example, a 5% increment in the real part of $L_{ij}$ is considered, and the predicted change in $\Delta \lambda_0$ can be calculated based on (26).

$$\Delta \lambda_0 = \left[ \boldsymbol{PF}_{\text{mod}}^Y \right]_{ji} \cdot \text{Re}\left( \Delta L_{ij} \right) \cdot 5\% \qquad (26)$$

For comparison, the system $\lambda_0$ is re-computed for 5% increment in the real part of $L_{ij}$ and the actual change from the original condition is denoted as $\Delta \lambda_0'$. The error between the predicted $\Delta \lambda_0$ and actual $\Delta \lambda_0'$ are calculated in Fig. 6(a), most of which are within 10%. The predictions are not perfect because steady-state operation points are assumed unchanged.

*2) Validation of Component and Parameter Sensitivity*

The same procedure is repeated for the error analysis of the component and parameter sensitivity. For example, a 5% increment in the real part of $Z_{\text{net},ij}$ and $\rho_z$ is considered, and the predicted change $\Delta \lambda_0$ can be calculated based on (27) and (28).

TABLE II
NOTATIONS OF COMPONENTS AND PARAMETERS

|  | Components | Parameters |
|---|---|---|
| $\boldsymbol{Y}_{\text{con1}}$ | $Y_{\text{con},11}$, $Y_{\text{con},12}$, $Y_{\text{con},13}$, $Y_{\text{con},21}$, $Y_{\text{con},22}$, $Y_{\text{con},23}$, $Y_{\text{con},31}$, $Y_{\text{con},32}$, $Y_{\text{con},33}$ | control bandwidth: $\alpha_{pll}^{S1}$ (PLL), $\alpha_{cc}^{S1}$ (inner current loop), $\alpha_{PQ}^{S1}$ (power loop) |
| $\boldsymbol{Y}_{\text{con2}}$ | $Y_{\text{con},44}$, $Y_{\text{con},45}$, $Y_{\text{con},46}$, $Y_{\text{con},54}$, $Y_{\text{con},55}$, $Y_{\text{con},56}$, $Y_{\text{con},64}$, $Y_{\text{con},65}$, $Y_{\text{con},66}$ | control bandwidth: $\alpha_{pll}^{S2}$ (PLL), $\alpha_{cc}^{S2}$ (inner current loop), $\alpha_{PQ}^{S2}$ (power loop) |
| $\boldsymbol{Y}_{\text{con3}}$ | $Y_{\text{con},77}$, $Y_{\text{con},78}$, $Y_{\text{con},79}$, $Y_{\text{con},87}$, $Y_{\text{con},88}$, $Y_{\text{con},89}$, $Y_{\text{con},97}$, $Y_{\text{con},98}$, $Y_{\text{con},99}$ | control bandwidth: $\alpha_{pll}^{R}$ (PLL), $\alpha_{cc}^{R}$ (inner current loop), $\alpha_{dc}^{R}$ (dc voltage loop), $\alpha_{PQ}^{R}$ (power loop) |
| $\boldsymbol{Z}_{\text{net1}}$ | $Z_{\text{net},11}$, $Z_{\text{net},12}$, $Z_{\text{net},21}$, $Z_{\text{net},22}$ | resistance of AC-grid 1: $R_{\text{ac}}^{g1}$ / reactance of AC-grid 1: $L_{\text{ac}}^{g1}$ |
| $\boldsymbol{Z}_{\text{net2}}$ | $Z_{\text{net},44}$, $Z_{\text{net},45}$, $Z_{\text{net},54}$, $Z_{\text{net},55}$ | resistance of AC-grid 2: $R_{\text{ac}}^{g2}$ / reactance of AC-grid 2: $L_{\text{ac}}^{g2}$ |
| $\boldsymbol{Z}_{\text{net3}}$ | $Z_{\text{net},77}$, $Z_{\text{net},78}$, $Z_{\text{net},87}$, $Z_{\text{net},88}$ | resistance of AC-grid 3: $R_{\text{ac}}^{g3}$ / reactance of AC-grid 3: $L_{\text{ac}}^{g3}$ |
| $\boldsymbol{Z}_{\text{net4}}$ | $Z_{\text{net},33}$, $Z_{\text{net},36}$, $Z_{\text{net},39}$, $Z_{\text{net},63}$, $Z_{\text{net},66}$, $Z_{\text{net},69}$, $Z_{\text{net},93}$, $Z_{\text{net},96}$, $Z_{\text{net},99}$ | dc capacitor: $C_{\text{dc}}^{S1}$, $C_{\text{dc}}^{S2}$, $C_{\text{dc}}^{R}$ / dc line resistance: $R_{\text{dc}}^{l1}$, $R_{\text{dc}}^{l2}$, $R_{\text{dc}}^{l3}$ / dc line reactance: $L_{\text{dc}}^{l1}$, $L_{\text{dc}}^{l2}$, $L_{\text{dc}}^{l3}$ |

TABLE III
INFORMATION OF EIGENVALUE DECOMPOSITION

| Critical Modal Loop Gain | | |
|---|---|---|
| $\Lambda_1^L$ : 5.6062-$j$2.1130 | $\Lambda_2^L$ : -1×10$^{-11}$-$j$1×10$^{-12}$ | $\Lambda_3^L$ : 0.2106-$j$0.0377 |
| $\Lambda_4^L$ : 1.2641+$j$0.1369 | $\Lambda_5^L$ : 1.1973+$j$0.1209 | $\Lambda_6^L$ : 0.8912-$j$0.0461 |
| $\Lambda_7^L$ : 0.9731-$j$0.0169 | $\Lambda_8^L$ : 1.0005+$j$0.0004 | $\Lambda_9^L$ : 1.0001+$j$0.0001 |
| Modification Factor $\xi^L$ : | | |
| $g^L$ | | 0.0473+$j$0.0694 |
| $\Lambda_1^L \Lambda_3^L \Lambda_4^L \Lambda_5^L \Lambda_6^L \Lambda_7^L \Lambda_8^L \Lambda_9^L$ | | 1.7027-$j$0.0740 |
| $\xi^L$ | | -10.6915+$j$17.2608 |

This is also valid for $Y_{con,ij}$ and $\rho_y$.

$$\Delta\lambda_0 = \left[\bm{Y}_{con}\bm{PF}_{mod}^L\right]_{ji} \cdot \text{Re}\left(\Delta Z_{net,ij}\right) \cdot 5\% \quad (27)$$

$$\Delta\lambda_0 = \sum\left(\left[\bm{Y}_{con}\bm{PF}_{mod}^L\right]_{ji} \frac{\partial Z_{net,ij}}{\partial \rho_z}\right) \cdot \left(\Delta\rho_z\right) \cdot 5\% \quad (28)$$

Similarly, the system $\lambda_0$ is re-computed for the above increment and the obtained actual $\Delta\lambda_0'$ are compared with predicted $\Delta\lambda_0$. The results of the corresponding error analysis are presented in Fig. 6(b), (c) and (e), where most errors are within 5%.

So far, the correctness of the proposed multilayer LGES analysis method has been validated, including indices of LG-PF, component- and parameter- level sensitivity.

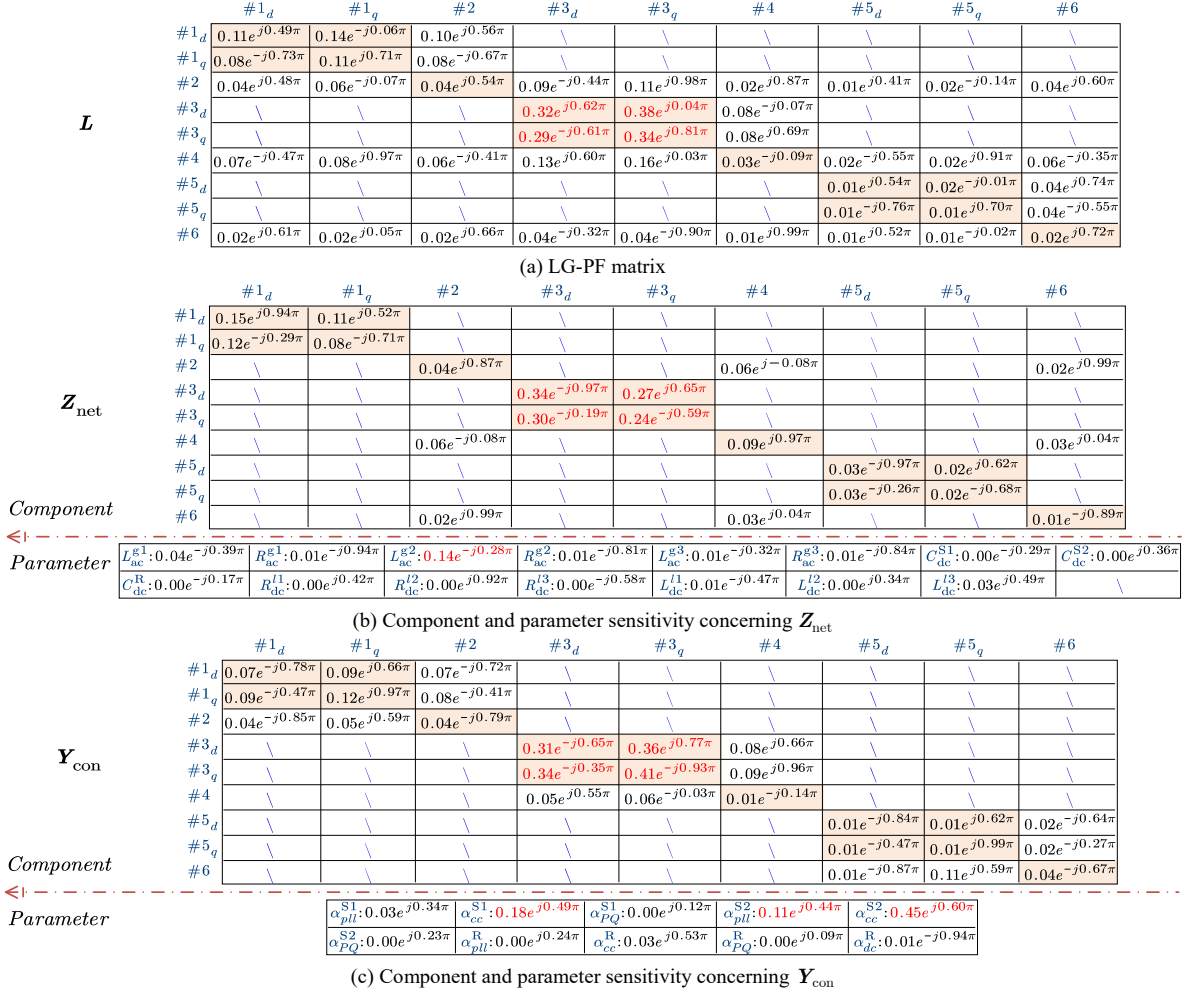

Fig. 5. Results of the multilayer LGES analysis on the test case.

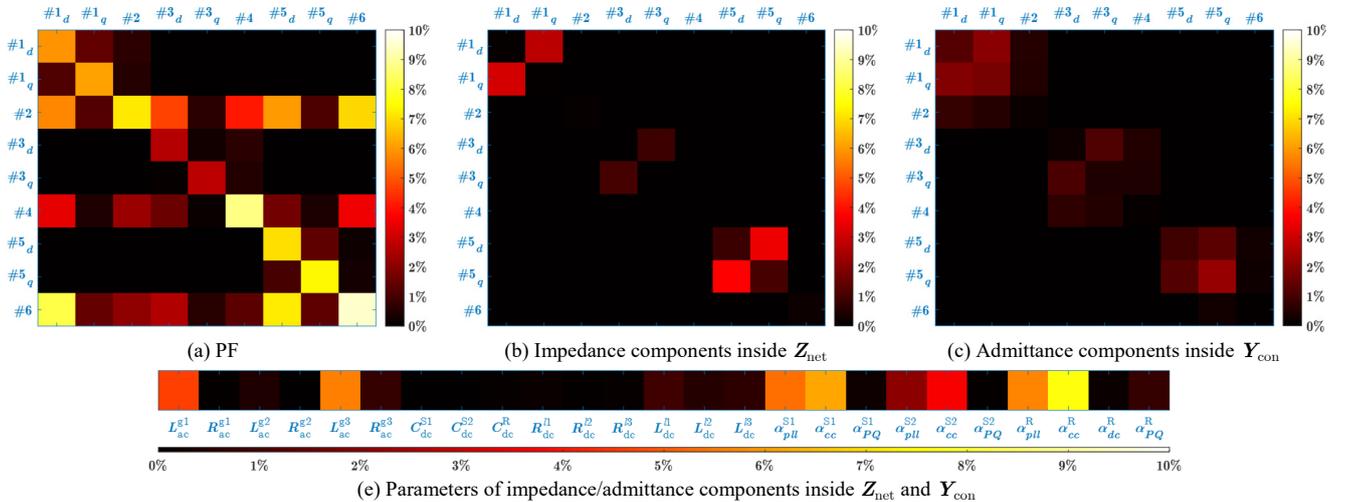

Fig. 6. Error Analysis of multilayer LGES indices (the magnitude error between theoretically predicted and numeric obtained sensitivities).

## V. Application on the Oscillation Suppression

This section will exhibit the potential of the proposed LGES method on the oscillation suppression, mainly the role of the parameter sensitivity played in guiding the controller tuning.

### A. Parameter Sensitivity-Guided Controller Tuning

As shown in Fig. 5 (b) and (c), four parameters $L_{ac}^{g2}$, $\alpha_{cc}^{S1}$, $\alpha_{pll}^{S2}$, $\alpha_{cc}^{S2}$ with high magnitude are regarded as dominant parameters in this oscillation case. Their parameter sensitivity given in Fig. 5 (b) and (c) are also depicted in the complex plane, as shown in Fig. 7(a). The length of the parameter sensitivity vector is exactly its magnitude, while the direction of the parameter sensitivity vector is namely its phase. If the parameter sensitivity vector is directed to the RHP unstable plane, the value of this parameter should be decreased, and vice versa. Such complex-plane characterization can be regarded as the traction action of parameters to the oscillation mode (e.g., $\lambda_1$ in the above case study).

As a comparison, the parameter sensitivity with respect to the critical modal loop gain (i.e., $\Lambda_2^L$ in this case) are also shown in Fig. 7(b). It can be obtained conveniently by replacing $\boldsymbol{PF}_{\text{mod}}^L$ in (22) and (24) with $\boldsymbol{PF}^L$, which is like:

$$\frac{\partial \Lambda_2^L}{\partial \rho_z} = \sum \left( \left[ \boldsymbol{Y}_{\text{con}} \boldsymbol{PF}^L \right]_{ji} \frac{\partial Z_{\text{net},ij}}{\partial \rho_z} \right) \quad (29)$$

$$\frac{\partial \Lambda_2^L}{\partial \rho_y} = \sum \left( \left[ \boldsymbol{PF}^L \boldsymbol{Z}_{\text{net}} \right]_{ji} \frac{\partial Y_{\text{con},ij}}{\partial \rho_y} \right) \quad (30)$$

Time-domain simulations are given in Fig. 7(c) to test which parameter sensitivity is more suitable for guiding the oscillation suppression. According to Fig. 7(a), the system will become more stable by decreasing $L_{ac}^{g2}$, and the oscillation frequency should increase. This is validated in Fig. 7(c) that the system becomes stable when decreasing $L_{ac}^{g2}$ by 3%, and the oscillation frequency increases from 8.5 Hz to 8.625 Hz. However, Fig. 7(b) gives a contrary predication that the system will become more stable by increasing $L_{ac}^{g2}$. Results of other parameters shown in Fig. 7(c) also conform with predictions in Fig. 7(a), for instance, the system becomes more stable if decreasing $\alpha_{cc}^{S1}$ and $\alpha_{pll}^{S2}$, or increasing $\alpha_{cc}^{S2}$. But Fig. 7(b) will give misleading information, and thus the established parameter sensitivity with respect to the oscillation mode is more suitable for guiding the controller tuning-based oscillation suppression.

### B. Discussion on the Reference for Determining Dominant Parameters

An interesting phenomenon can be observed by comparing Fig. 5(b), (c) with Fig. 7(c) that the magnitude of the parameter sensitivity may not accurately reflect the dominance of parameters on the stability margin. In Fig. 7(c), all parameter changes are under the same ratio, i.e., 3%, so as to compare suppression effects of each parameter fairly. It can be seen that suppression effects of $\alpha_{cc}^{S2}$ and $L_{ac}^{g2}$ are the most obvious. This is reasonable for $\alpha_{cc}^{S2}$, whose sensitivity magnitude is the largest as 0.45 given in Fig. 5(c). However, the sensitivity magnitude of $\alpha_{cc}^{S1}$ is greater than that of $L_{ac}^{g2}$ (0.18>0.14 shown in Fig. 5(c)),

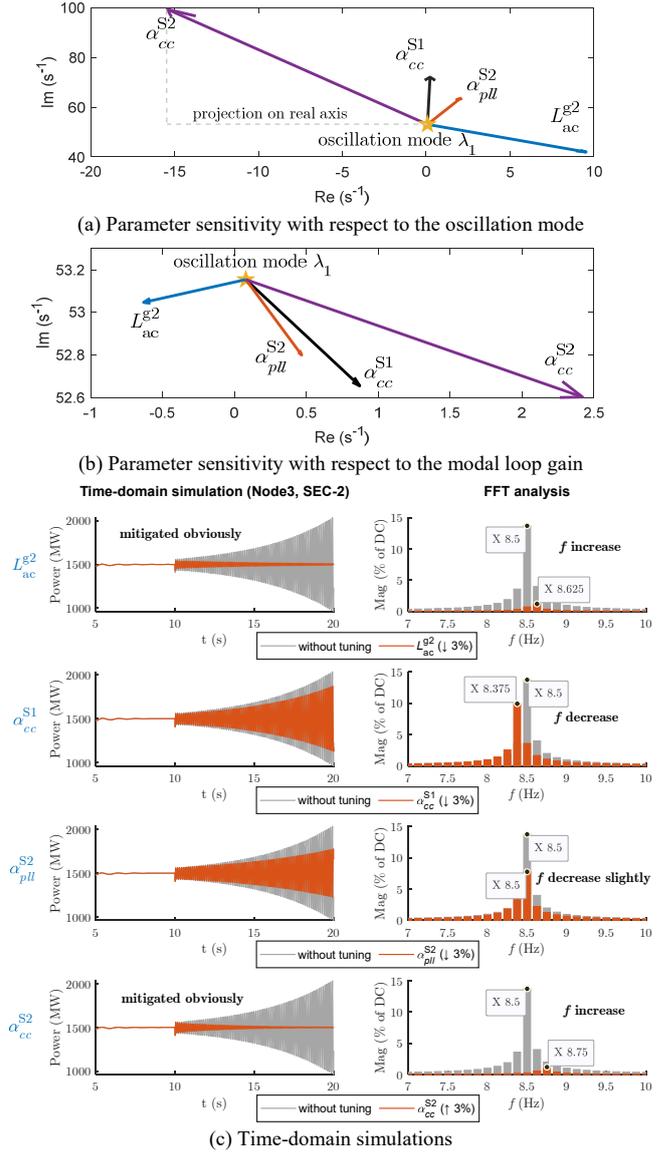

Fig. 7. Parameter sensitivity-based controller tuning.

while suppression effect of $\alpha_{cc}^{S1}$ is far weaker than $L_{ac}^{g2}$.

The reason behind is that the system stability margin is more related to the movement of the oscillation mode on the real axis, while the magnitude of the parameter sensitivity is a comprehensive evaluation for the influence of both real axis and imaginary axis. It can be seen in Fig. 7(a) that the sensitivity phase of $\alpha_{cc}^{S1}$ is nearly $0.5\pi$ (almost perpendicular to the real axis), meaning that its influence imposed on the real axis is small despite of large magnitude.

The comparison on the magnitude of different parameter sensitivities is presented in Fig. 8. Para-Sen1 denotes the magnitude of the parameter sensitivity to the modal loop gain (i.e., traditional LGES) shown in (29) and (30). Para-Sen2 denotes the magnitude of the proposed parameter sensitivity to the oscillation mode, and its projection on real axis is denoted as Para-Sen3. The per unit transformation is applied to the Para-Sen1, Para-Sen2 and Para-Sen3. It can be seen that Para-Sen1 is the same as Para-Sen2, because constant $\xi^L$ will be offset in

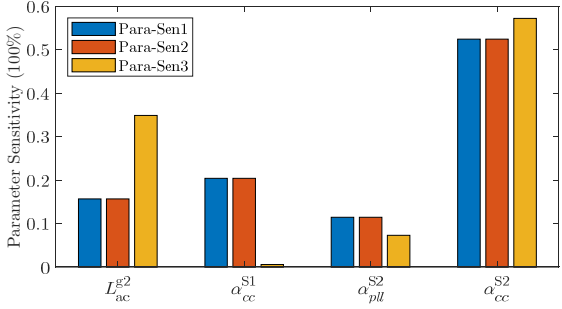
Fig. 8. Comparison on the magnitude of different parameter sensitivities.

the per unit transformation. Taking $\rho_1$ for example:

$$\overline{\frac{\partial \lambda_1}{\partial \rho_1}} = \frac{\left|\frac{\partial \lambda_1}{\partial \rho_1}\right|}{\sum_{i=1}^{n}\left|\frac{\partial \lambda_1}{\partial \rho_i}\right|} = \frac{\left|\xi^L \frac{\partial \Lambda_2^L}{\partial \rho_1}\right|}{\sum_{k=1}^{n}\left|\xi^L \frac{\partial \Lambda_2^L}{\partial \rho_i}\right|} = \frac{\left|\xi^L\right|\left|\frac{\partial \Lambda_2^L}{\partial \rho_1}\right|}{\sum_{k=1}^{n}\left|\xi^L\right|\left|\frac{\partial \Lambda_2^L}{\partial \rho_i}\right|} = \overline{\frac{\partial \Lambda_2^L}{\partial \rho_1}}$$

(31)

According to Para-Sen3 in Fig. 8, the dominance order should be as: $\alpha_{cc}^{S2} > L_{ac}^{g2} > \alpha_{pll}^{S2} > \alpha_{cc}^{S1}$, which fits well with time-domain simulations in Fig. 7(c). Therefore, the projection of the established parameter sensitivity on the real axis is more suitable as the reference for determining dominant parameters, in terms of the oscillation suppression.

## VI. CONCLUSION

This paper proposes a multilayer LGES analysis method to facilitate the oscillation source locating and suppression of the converter-dominated system. Three crucial indices directed to the time-domain oscillation mode are established, including loop gain-based PF, component- and parameter- sensitivity, whose correctness is validated. Main conclusions include:
1) Traditional LGES is in essence a frequency-domain modal characteristic, which does not necessarily point to the stability (or oscillation mode) as that of the time-domain eigen-sensitivity. Thus, the derived parameter sensitivity cannot provide an effective guidance for the suppression on the unstable oscillation mode;
2) A complex-valued constant transformation is proposed to link the traditional LGES with the time-domain oscillation mode. In this way, time-domain sensitivity information originally defined on the state space model can now be acquired using the more modularized frequency-domain model, e.g., loop gain model;
3) The established parameter sensitivity contains the information of both magnitude (absolute values) and phase (direction). Its projection on the real axis is more suitable for determining dominant parameters in terms of the oscillation suppression.

On the foundation of the established LGES method, its connection with another widely-applied eigen-sensitivity method, i.e., node admittance-based eigen-sensitivity method (NAES), will be further studied in the future work, including their similarities and differences on the physical meaning, oscillation suppression, etc.